\documentclass[%
 reprint,
 amsmath,amssymb,
 aps,prl
]{revtex4-2}

\usepackage{graphicx}
\usepackage{dcolumn}
\usepackage{bm}
\usepackage{braket}
\usepackage{mathrsfs}
\usepackage{lipsum}
\usepackage{mathtools}
\usepackage{siunitx}
\usepackage{nicefrac}

\begin{document}

\title{Reversal of quantised Hall drifts at non-interacting and interacting topological boundaries}

\author{Zijie Zhu}
\email[These authors contributed equally.]{}
\author{Marius Gächter}
\email[These authors contributed equally.]{}
\author{Anne-Sophie Walter}
\author{Konrad Viebahn}
\email[]{viebahnk@phys.ethz.ch}
\author{Tilman Esslinger}
\affiliation{Institute for Quantum Electronics \& Quantum Center, ETH Zurich, 8093 Zurich, Switzerland
}

\begin{abstract}
The transport properties of gapless edge modes at boundaries between topologically distinct domains are of fundamental and technological importance.
Therefore, it is crucial to gain a better understanding of topological edge states and their response to interparticle interactions.
Here, we experimentally study long-distance quantised Hall drifts in a harmonically confined topological pump of non-interacting and interacting ultracold fermionic atoms.
We find that quantised drifts halt and reverse their direction when the atoms reach a critical slope of the confining potential, revealing the presence of a topological boundary.
The drift reversal corresponds to a band transfer between a band with Chern number $C = +1$ and a band with $C = -1$ via a gapless edge mode, in agreement with the bulk-edge correspondence for non-interacting particles.
We establish that a non-zero repulsive Hubbard interaction leads to the emergence of an additional edge in the system, relying on a purely interaction-induced mechanism, in which pairs of fermions are split.
\end{abstract}

\maketitle

\paragraph{}
The existence of individual edge modes at topological boundaries plays a crucial role in quantum Hall physics.
More specifically, a non-trivial topology in the bulk of a material ensures that its edge modes are gapless and chiral.
Gaplessness is related to the bulk-edge correspondence, stating that the number of topological edge modes is equal to the difference in Chern number across an interface~\cite{%
hasan_colloquium_2010}.
Consequently, a gapless mode should allow an adiabatic transfer from one band to another, resulting in a reflection of transverse bulk currents in the opposite direction if the two bands feature opposite Chern numbers.
However, the coherence time in most electronic materials is not sufficient to observe this effect, and edges are generally probed spectroscopically~\cite{hasan_colloquium_2010,hafezi_imaging_2013,yatsugi_observation_2022,xiang_simulating_2023}.
Moreover, studies of edge physics in engineered quantum systems, such as ultracold atoms and photonics, have so far been focussed on chirality~\cite{atala_observation_2014,stuhl_visualizing_2015,mancini_observation_2015,chalopin_probing_2020,ozawa_topological_2019} and localisation~\cite{kraus_topological_2012,leder_real-space_2016,meier_observation_2016,an_direct_2017}.
A boundary reflection has not been detected~\cite{jotzu_experimental_2014,aidelsburger_measuring_2015,nakajima_topological_2016,lohse_thouless_2016,minguzzi_topological_2022}, or it was disregarded~\cite{qian_quantum_2011,bellec_non-diffracting_2017}, and to our knowledge it has never been studied for variable interaction strength.
Here, we observe the reversal of quantised bulk drifts due to harmonic trapping in a topological Thouless pump, the temporal analogue of the quantum Hall effect~\cite{cooper_topological_2019,oka_floquet_2019,citro_thouless_2023}.
The reflection is a fundamental manifestation of confined topological matter and directly shows the gapless nature of topological edge modes.
Going beyond the non-interacting regime, we discover the emergence of a second edge for repulsive Hubbard $U$.

\begin{figure}
    \centering
    \includegraphics[width = 0.5\textwidth]{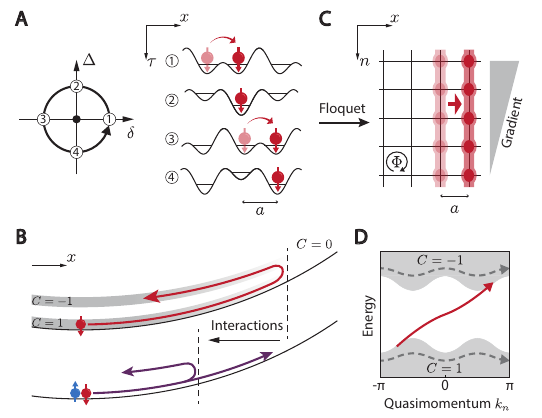}
    \caption{\textbf{Reflection of quantised Hall drifts off a topological interface.} (\textbf{A}) Topological trajectory in the Rice-Mele model enclosing the critical point (black dot), which is the origin in the parameter space spanned by bond dimerisation $\delta$ and superlattice offset $\Delta$. (\textbf{B}) Topological pump in the presence of confining potentials. In the non-interacting case (top), a harmonic trap gives rise to topological trivial ($C=0$) and non-trivial ($C\neq 0$) regions, separated by a topological interface. The atoms exhibit a quantised drift until they are reflected at the interface. With repulsive on-site interactions (bottom), the reflection happens closer to the centre, accompanied by atoms still drifting in the original direction. (\textbf{C}) Using Floquet theory, the 1D Rice-Mele pump can be mapped to a 2D Harper-Hofstadter-Hatsugai model with a linear gradient along the synthetic dimension $n$ which represents the photon number. The magnetic flux per plaquette is $\Phi=1/2$ in units of the magnetic flux quantum~\cite{Note1}. The gradient along $n$ leads to a transverse Hall drift along $x$ (red arrows) due to the nontrivial topology of the bands. (\textbf{D}) Schematic spectrum of the mapped 2D Hofstadter model in a semi-infinite geometry. The lowest two bands have $C=\pm1$, respectively. The linear gradient induces Bloch oscillations in the synthetic reciprocal space (dashed arrows). A gapless edge mode (solid arrow) appears at the topological interface. The reflection of the Hall drift can be understood as atoms being transported from the lower band ($C=1$) to the higher band ($C=-1$) via the topological edge mode.}
    \label{fig:1}
\end{figure}

\paragraph{}
The experiments are performed with ultracold fermionic potassium-40 atoms, which are loaded into a three-dimensional optical lattice of wavelength $\lambda = \SI{1064}{nm}$~\cite{walter_quantization_2023}.
The optical potential along the $x$-direction exhibits short ($a = \lambda/2$) and long ($2a$) periodicities, whose relative phase $\varphi$ is dynamically tuneable, while the transverse directions ($y$ and $z$) are effectively frozen out.
We initiate the topological pump by linearly ramping $\varphi$ in time, which cyclicly modulates the lattice potential with a period $T$, as shown in Fig.~\ref{fig:1}A.
In response to the adiabatic modulation, the atoms exhibit a quantised drift of one unit cell per pump cycle.
In contrast to previous experimental realisations with bichromatic lattices~\cite{lohse_thouless_2016,nakajima_topological_2016,nakajima_competition_2021}, our setup employs a robust single-wavelength design and topological pumping persists for more than a hundred cycles.

We describe our system using the periodically modulated Rice-Mele-Hubbard Hamiltonian with harmonic confinement,
\begin{eqnarray}\label{eqn:RM}
    \hat{H}(\tau) &=& - \sum_{j,\sigma}\left[t + (-1)^j\delta(\tau)\right]\left(\hat{c}_{j\sigma}^\dagger \hat{c}_{j+1\sigma} + \text{h.c.}\right) \\
    \nonumber &&+\,\Delta(\tau)\sum_{j,\sigma} (-1)^j \hat{c}_{j\sigma}^\dagger \hat{c}_{j\sigma}+U\sum_{j} \hat{c}_{j\uparrow}^\dagger \hat{c}_{j\uparrow}\hat{c}_{j\downarrow}^\dagger \hat{c}_{j\downarrow}~\\
    \nonumber &&+\,\sum_{j,\sigma} V_{j}\hat{c}_{j\sigma}^\dagger \hat{c}_{j\sigma}~,
\end{eqnarray}
where $\hat{c}_{j\sigma}$ is the fermionic annihilation operator for spin $\sigma \in \{\uparrow,\downarrow \}$ on site $j$, and $t$ denotes the average tunnelling.
The periodic modulation is characterised by the parameters $\delta(\tau)=\delta_{0}\cos(2\pi \tau/T)$ (bond dimerisation) and $\Delta(\tau)=\Delta_{0}\sin(2\pi \tau/T)$ (sublattice offset).
The topological nature of the pump becomes evident when considering the trajectories in the $\delta$--$\Delta$ plane.
If the trajectory encloses the origin, referred to as critical point (Fig.~\ref{fig:1}A), it causes the quantised drift motion~\cite{citro_thouless_2023,cooper_topological_2019,wang_topological_2013}.

\paragraph{}
The harmonic confinement is characterised by the trap frequency $\nu$, entering Eq.~\ref{eqn:RM} as $V_{j} = \frac{1}{2}m(2\pi\nu a j)^2 \equiv V_{0}j^{2}$ ($m$ is the atomic mass).
Due to the confinement, the atoms are initially located at the centre of the trap.
Topological pumping then leads to a quantised drift of atoms against the confining potential (Fig.~\ref{fig:1}B).
Our measurements show that the quantised drift changes its direction at a certain distance from the trap centre.
We will demonstrate that this happens when the gradient of the confinement overcomes the band gap and a boundary between topological and trivial regions emerges.
For repulsive interactions, we observe another reflection, closer to the trap centre, while a part of the atoms keeps  drifting in the original direction (Fig.~\ref{fig:1}B).  

\paragraph{}In the following, we develop a description of the reflection in terms of gapless edge modes 
and the bulk-edge correspondence within the framework of the  Harper-Hofstadter-Hatsugai (HHH) model with one real ($x$) and one synthetic ($n$) dimension.
The model features bulk Chern bands with $C = +1$ and $C = -1$.
An exact mapping between the non-interacting 1D Rice-Mele Hamiltonian (Eq. 1) and the two-dimensional (2D) HHH model can be obtained using Floquet theory, illustrated in Fig.~\ref{fig:1}C (for derivation see, e.g., refs.~\cite{oka_floquet_2019} and~\footnote{see materials and methods}).
A linear gradient along the synthetic dimension $n$ appears in the mapping since the state with $n$ photons acquires an energy of $-n\hbar\omega$, where $\omega=2\pi/T$ is the pump frequency.
The gradient along $n$ or, equivalently, an external force causes Bloch oscillations along the synthetic reciprocal dimension $k_{n}$ which, in turn, lead to a Hall drift or `anomalous velocity' along the transverse real direction $x$~\cite{jotzu_experimental_2014,aidelsburger_measuring_2015}.
The bulk Hall drift along $x$ corresponds exactly to the quantised displacement measured in the topological pump.
The trap induces a boundary between topological ($C_{\text{centre}} = 1$) and trivial ($C_{\text{right}} = 0$) regions and a single gapless edge mode emerges, according to the bulk-edge correspondence: $C_{\text{centre}} - C_{\text{right}} = 1$.
The edge modes connects two bands of opposite Chern invariant, as shown in  Fig.~\ref{fig:1}D.
Thus, a Bloch oscillation transfers the atoms from the ground to the first excited band via that edge mode.
Since the first excited band has Chern number $-1$ the atoms are now moving `backwards', resulting in a reversal of the quantised Hall drift.

\begin{figure*}[t!]
    \centering
    \includegraphics[width = 0.48\textwidth]{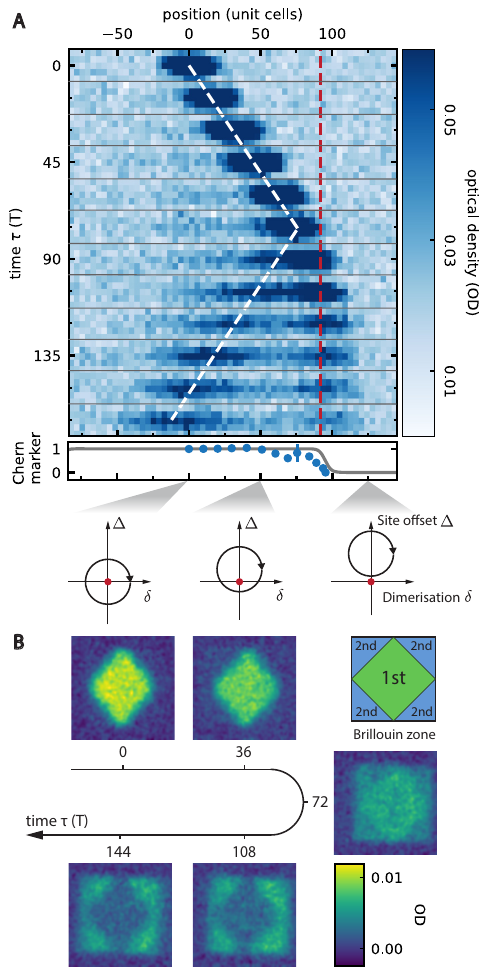}
    \caption{\textbf{Measuring the reversal of a quantised Hall drift.} (\textbf{A}) The atomic in-situ images at timesteps $\tau /T = 0, 15, 30, \dots$ show a quantised drift along $x$ before the atoms are reflected at the topological boundary.
    Each density image (60 by 5 pixels) is averaged over three individual measurements with the parameters $V_{0}=0.0149(9)t$, $\Delta_{0}=2.7(1)t$, and $T=\SI{3}{ms}=12.8(2)\hbar/t$.
    The red dashed line indicates the topological boundary $x_{\rm{edge}}/\left(2a\right)\approx\frac{1}{2}\Delta_{0}/V_{0}$.
    The white dashed lines are linear fits to the atom drift, yielding slopes of $1.00(1)\times 2a/T$ before, and $-0.99(3)\times 2a/T$ after the reflection. Cloud positions, averaged over the transverse direction, are fitted using Gaussians.
    The experimental Chern marker (lower panel, points) is determined by the velocity of the right-moving cloud at different positions.
    The theoretical Chern marker (line) is calculated in a phenomenological model with $V_{j}=V_{0}\left(-1\right)^{j}j$ which has the same absolute value of the local tilt $\Delta_{\rm{ext}}(j)=V_{0}\left|j-\frac{1}{2}\right|$ as the harmonic trap~\cite{Note1}.
    In a local density approximation picture, the local tilt $\Delta_{\rm{ext}}$ shifts the $\delta$--$\Delta$ pump trajectory upwards~\cite{nakajima_topological_2016}.
    Depending on whether or not the trajectory encloses the critical point, the pump is rendered topological or trivial.
    (\textbf{B}) Measured band populations as a function of time $\tau$. Each density image is averaged over six individual measurements with the parameters $V_{0}=0.0199(6)t'$, $\Delta_{0}=3.3(2)t'$ and $T=\SI{3}{ms}=10.3(2)\hbar/t'$~\cite{Note1}. 
    The total atom number remains constant, within error bars, throughout the experiment.
    Due to the underlying honeycomb lattice geometry in the $x$--$z$ plane, the first Brillouin zone has a diamond shape.
            The band population is inverted when the bulk current is reflected off the topological interface, manifesting the gapless nature of the topological edge mode.}
    \label{fig:2}
\end{figure*}

\paragraph{}Figure~\ref{fig:2}A shows experimental in-situ images of the atomic cloud as a function of time $\tau$ at $U = 0$.
The data shows a quantised drift of $1.00(1)\times 2a/T$ up to about $60\,T$, which confirms the long coherence time of Bloch oscillations which induce the transverse drift.
At $\tau \simeq 75\;T$ the atoms change their drift direction, which is a key observation of this work.
The expected topological boundary (red dashed line) represents the position at which the local tilt from the external harmonic potential $\Delta_{\rm{ext}}\left(j\right)\equiv\frac{1}{2}\left|V_{j}-V_{j-1}\right|=V_{0}\left|j-\frac{1}{2}\right|$ equals the maximum sublattice offset $\Delta_{0}$, thus,  $x_{\rm{edge}}/\left(2a\right)\simeq\frac{1}{2}\Delta_{0}/V_{0}=92(7)$. 
Beyond this position the total sublattice offset ceases to change sign, rendering the region outside $x_{\rm{edge}}$ topologically trivial.
The boundary caused by the harmonic confinement is not infinitely sharp, but smoothened over several lattice sites.
This leads to a small $T$--dependence of the reflection position (Fig.~\ref{fig:si_vs}), compared to its absolute value, and the calculation above should be understood as the outermost point of the reflective region.
The reflected atoms exhibit a quantised drift of $-0.99(3)\times 2a/T$ in the opposite direction, in agreement with a transfer to the first excited band with $C=-1$.
The linear relation between the position of topological boundary $x_{\rm{edge}}$ and the maximum sublattice offset $\Delta_{0}$ is further confirmed by measuring the reflection in different lattices (Fig.~\ref{fig:si_vs}). 
The reflection is observed under all parameter settings tested in this work, highlighting that the existence of the topological boundary is robust.
In addition to the reflection, we observe a cloud of atoms temporarily residing at the turning point before being reflected.
We confirm this with numerical simulations which show that eventually all atoms drift back (Fig.~\ref{fig:si_turnaround_spectrum}) and we find agreement with the experimentally detected retention timescale (Fig.~\ref{fig:si_retention}). 
The dynamics at the smooth boundary between topologically trivial and nontrivial regions is discussed in more detail in the Materials and methods.

\begin{figure*}[t!]
    \centering
    \includegraphics[width = 1\textwidth]{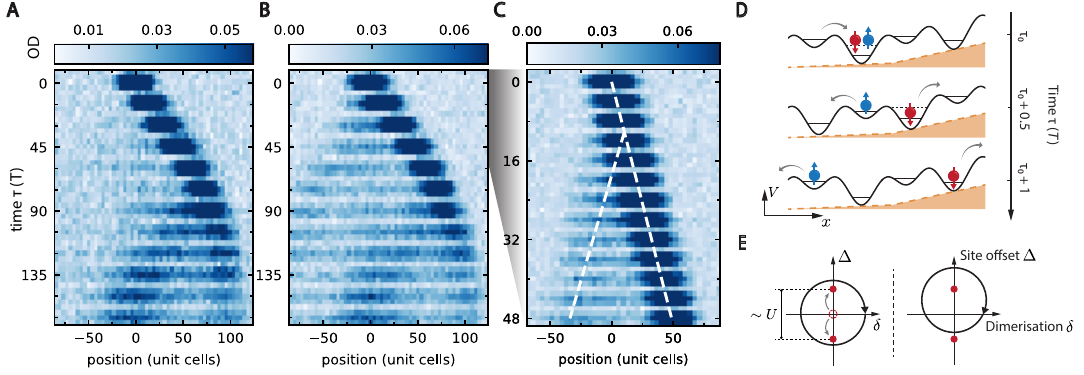}
    \caption{\textbf{Reflection of quantised Hall drifts from an interacting topological edge.} (\textbf{A}) An attractive Hubbard interaction of $U=-3.48(7)t$ leads to the same reflection behaviour as observed for $U= 0$ (measurement parameters otherwise identical to Fig.~\ref{fig:2}A).
    (\textbf{B}) Repulsive Hubbard interactions of $U=3.48(7)t$ lead to the emergence of a second reflection, closer to the trap centre, which we attribute to an interacting topological boundary.
    A zoom-in (\textbf{C}) shows that the early reflection happens after about twelve cycles. The white dashed lines are guides to the eye, calculated as linear fits to the cloud position, extracted as the sum of a skewed and a regular Gaussian.
    (\textbf{D}) Microscopic description of the interaction-induced reflection for repulsive Hubbard $U$.
    When the maximum energy offset between two neighbouring sites $2\left(\Delta_{0}-\Delta_{\rm{ext}}\right)$ becomes smaller than the Hubbard interaction, formation of double occupancies is prohibited and one atom is left in the higher-energy site of a unit cell, which then drifts in the opposite direction.
    (\textbf{E}) The critical point in the $\delta$--$\Delta$ plane is split into two in the presence of repulsive Hubbard $U$~\cite{bertok_splitting_2022}. When the pump trajectory encloses both critical points, a quantised drift is expected, as in the non-interacting system. The local tilt given by the external potential $\Delta_{\rm{ext}}$ shifts the trajectory along the $\Delta$--axis, eventually enclosing just one critical point.
    Thus, the value of the topological invariant (see main text) changes in space, leading to the emergence of an interaction-induced boundary.
    }
    \label{fig:3}
\end{figure*}

\paragraph{} We support the in-situ images with measurements of band population before, during, and after the reflection, as shown in Fig.~\ref{fig:2}B~\cite{Note1}.
Before the reflection, we find a uniformly occupied ground band, which is consistent with the observation of a quantised Hall drift. 
At the reflection ($\tau\simeq 72\;T$), we observe an inversion of the population to the first excited band.
After the reflection, the inverted population remains almost unchanged while the atoms are travelling back, highlighting the absence of incoherent relaxation to the ground band, even after more than a hundred Bloch oscillations.

\paragraph{}We further explore the effect of attractive and repulsive interactions on the topological boundary.
For attractive Hubbard $U=-3.48(7)t=-1.27(7)\Delta_{0}$ (Fig.~\ref{fig:3}A), the quantised Hall drift is reversed at the same position as in the non-interacting situation.
This can be explained in terms of the Rice-Mele model in which fermions in the strongly attractive regime approach the limit of hard-core bosons ~\cite{walter_quantization_2023}, and the condition for the emergence of a topological boundary $\Delta_{\rm{ext}}\left(j\right)=\Delta_{0}$ remains unchanged. 

\paragraph{}For repulsive Hubbard $U=3.48(7)t$ (Fig.~\ref{fig:3}B), we observe a second reflection in addition to the original one.
Compared to the non-interacting case, this reflection appears much closer to the trap centre. 
The zoomed-in image (Fig.~\ref{fig:3}C) shows that a proportion of the atoms start to move backwards after about 12 cycles.
In contrast to the drift reversal in the non-interacting system, a large fraction of the atom cloud still undergoes quantised drifting in the original direction; an upper bound for the amount of atoms reflected is given by the number of doubly occupied unit cells~\cite{Note1}.
In the following, we develop a microscopic picture of the interaction-induced partial reflection in the limiting case of two isolated spins $(\uparrow,\downarrow)$, which approximates our initial state in a unit cell (Fig.~\ref{fig:3}D). 
As long as the maximum energy offset between two neighbouring sites $2\left(\Delta_{0}-\Delta_{\rm{ext}}\right)$ is larger than the Hubbard $U$, the formation of a double occupancy is energetically allowed and the quantised drift persists~\cite{walter_quantization_2023}, even in the presence of an external potential.
However, when $\Delta_{\rm{ext}}$ becomes larger, the energy offset between two neighbouring sites remains always smaller than $U$ and double occupancy formation becomes prohibited. 
In the latter case, one atom of a singlet pair is transferred to the energetically excited site of a unit cell, which will subsequently drift in the opposite direction. 
The other atom, in the lower-energy site, will move onwards because on-site interactions become irrelevant if there is only one atom per unit cell.
Since the underlying Hamiltonian (Eq.~\ref{eqn:RM}) is $\text{SU}(2)$--symmetric, spin--$\uparrow$ and spin--$\downarrow$ have equal probability of being reflected and they remain correlated after the splitting process (Fig.~\ref{fig:si_singlet}).

\paragraph{}
The full many-body description of the interaction-induced reversal requires the development of suitable topological invariants 
for smooth confinements and strong interactions, which goes beyond the scope of this work.
Nevertheless, we obtain an intuition of the boundary's topological origins 
using again the idea of shifted pump trajectories in the $\delta$--$\Delta$ plane with a staggered potential (c.f.~Fig.~\ref{fig:2}A).
Numerical simulations have shown that a repulsive Hubbard $U$ can split the critical point at the origin into two separate ones~\cite{bertok_splitting_2022}.
The topological invariant associated with enclosing one of the two critical points is a $2\pi$ winding of the charge Berry phase, compared to a winding of $4\pi$ in the non-interacting case.
The distance between the new critical points is approximately $U$ up to a correction on the order of the tunnelling $t$~\cite{torio_phase_2001}.
When the trajectory encloses both critical points, quantised drift of two spins $(\uparrow,\downarrow)$ is expected, as in the non-interacting system. 
As the position-dependent local tilt $\Delta_{\rm{ext}}(j)$ shifts the trajectory upwards, it will enclose only one of the critical points beyond certain position along $x$ (Fig.~\ref{fig:3}E).
This implies a change of a topological invariant as function of space which, according to the bulk-edge correspondence, should lead to the emergence of an interacting topological edge.
The estimation of the interacting boundary at $\Delta_{\rm{ext}}(j)\simeq\Delta_{0}-U/2$ lies close to the centre and agrees with the microscopic picture discussed above.
Similar to the non-interacting case, this boundary should be considered as the outermost position of the reflective region.

\paragraph{}In conclusion, we have experimentally observed a reversal of quantised Hall drifts at a topological boundary in a harmonic potential. 
The reflection is a direct manifestation of the gapless nature of topological edge modes between Chern bands of opposite sign.
We explore the effect of Hubbard interactions, both attractive and repulsive, and find an asymmetric behavior with respect to $U=0$.
While on the attractive side the topological boundary is unaffected, repulsive interactions lead to the emergence of a second interface, featuring a splitting of quantised drifts.
As a result, our experiments could enable the realisation of circular current patterns for constructing novel many-body phases~\cite{letscher_growing_2015}.
More broadly, our work allows the exploration of the bulk-edge correspondence in the presence of interactions~\cite{irsigler_interacting_2019}, as well as the investigation of edge reconstruction~\cite{chklovskii_electrostatics_1992} in the quantum Hall effect and in interacting topological insulators.

\nocite{zhu_dataset_2023}
\nocite{uehlinger_artificial_2013,asboth_short_2016,stanescu_topological_2010,buchhold_effects_2012,kolovsky_quantum_2014,goldman_creating_2016,bianco_mapping_2011,mook_edge_2014}

\section*{Acknowledgments}
We would like to thank Jason Ho, Gian-Michele Graf, Thomas Ihn, Fabian Grusdt, Fabian Heidrich-Meisner, Armando Aligia, and Eric Bertok for valuable discussions. We also thank Julian Léonard and Nur Ünal for comments on an earlier version of the manuscript.
We would like to thank Alexander Frank for his contributions to the electronic part of the experimental setup.
We acknowledge funding by the Swiss National Science Foundation (Grant Nos.~182650, 212168, NCCR-QSIT, and TMAG-2 209376), Quantera dynamite PCI2022 132919, and European Research Council advanced grant TransQ (Grant No.~742579).


%

\clearpage
\newpage

\setcounter{figure}{0} 
\setcounter{equation}{0} 

\renewcommand\theequation{S\arabic{equation}} 
\renewcommand\thefigure{S\arabic{figure}} 
\renewcommand\thetable{S\arabic{table}} 

\setcounter{page}{1}

\onecolumngrid

\section{Materials and methods}

\subsection{Dependence of the drift reversal on experimental parameters}
\begin{figure}[htbp]
    \centering
    \includegraphics[width = 0.5\textwidth]{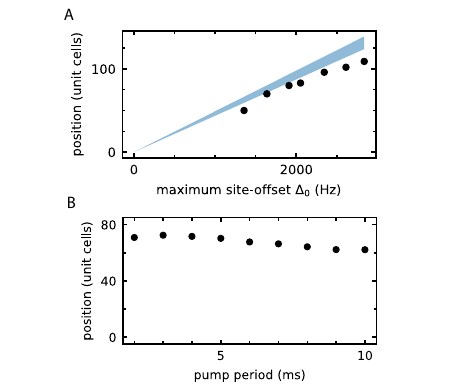}
    \caption{ \textbf{Experimental dependence of the reflection position.} The reflection point is expected to depend linearly on the maximal site-offset per pump cycle $\Delta _0$ which is experimentally verified in (\textbf{A}). Deviations for large values of $\Delta _0$ can be explained by the finite waist of the laser beams forming the lattice. 
    (\textbf{B}) shows the period dependence of the reflection position. Changing the pump period $T$ over an order of magnitude only changes the reflection point by about 10 unit cells, which is a result of the smooth boundary of a harmonic potential.    }
    \label{fig:si_vs}
\end{figure}

The expected drift reversal happens when the maximum local site offset over one pump-cycle $\Delta _0$ is equal to the local tilt given by the harmonic trap. This position given by $x_{\rm{edge}} \simeq \Delta _0 a / V_0$ with $a=\lambda /2$ and $V_0=\frac{1}{2}m(2\pi \nu a)^2$. By measuring the reflection point in lattices with different $\Delta_0$, we verify the relevant scaling $x_{\rm{edge}} \propto \Delta _0$, as shown in Fig. \ref{fig:si_vs}. The blue line in Fig. \ref{fig:si_vs}A marks the theoretically expected $x_{\rm{edge}}$ with the uncertainty propagated from the uncertainty of the trap frequency $\nu$. The disagreement between theory and experiment for larger values of $\Delta _0$ can be explained by the finite waist of the lattice beams.
In order to explore the edge in our system, the atoms are pumped by almost 100 unit cells ($\sim \SI{100}{\mu m}$).
Due to the Gaussian envelope of the transverse beams, which are essential to realise the pump, the lattice is effectively shallower far away from the centre.
Thus, $\Delta _0$ decreases towards the edge and atoms are reflected sooner.

We also find a small dependence of the reflection point on the pump period, compared to its absolute value, which spans roughly 10 unit cells when changing $T$ from $\SI{2}{ms}$ to $\SI{10}{ms}$ (Fig. \ref{fig:si_vs}B).
This can be understood by considering the energy spectrum of the Harper-Hofstadter-Hatsugai (HHH) model in a harmonic potential, which will be discussed below.

\subsection{Experimental sequence}
We start by preparing a degenerate cloud of fermionic $^{40}\rm{K}$ in a crossed dipole trap. 
We have a spin mixture of $m_F=\{-9/2,-7/2\}$ except for the measurements in Fig.~\ref{fig:2}B and Fig.~\ref{fig:si_vs}, where a spin-polarised cloud in the magnetic state $F=9/2,\ m_{F}=-9/2$ is used.
The spin-polarised cloud is directly loaded into the pumping lattice, while the spin mixture is first loaded into an intermediate chequerboard lattice with strongly attractive interactions. The two-step loading precludes the presence of atoms in the higher band and gives a larger fraction of atoms in doubly occupied unit cells. 
On attractive side we have around 80\% of atoms in doubly occupied unit cells and on repulsive side this value is about 50\%  ~\cite{walter_quantization_2023}.

After pumping the system for varying times, we either take a in-situ absorption image to measure the density or detect the band population with band-mapping. The latter is implemented with an exponential ramp to switch off the lattice beam in $\SI{500}{\mu s}$, followed by a time-of-flight expansion of $\SI{25}{ms}$ before absorption imaging.

For fitting the small reflected part in Fig.~\ref{fig:3}C the following function was used: \verb|scipy.stats.skewnorm.pdf|.

\subsection{Realisation of a Thouless pump in the Rice-Mele model}
The lattice setup is comprised of non-interfering standing waves in $x$, $y$, and $z$ directions, together with interfering laser beams in the $x$--$z$ plane. All the lattice beams come from a single laser source at wavelength $\lambda=\SI{1064}{nm}$.
These potential combine to form a honeycomb lattice in the $x$--$z$ plane, which can be considered as isolated tubes of one-dimensional superlattices along $x$ in the limit of deep transverse lattices. 
In each 1D tube the potential can be modeled by a one-dimensional superlattice with two sites per unit cell.
The total lattice potential, as seen by the atoms, is
\begin{equation}\label{eqn:potential}
\begin{split}
     &V(x,y,z,\tau) = \\
     &\quad- V_{\text{X}} I_{\text{self}} \cos^2 (kx + \vartheta / 2) \\
     &\quad- V_{\text{Xint}} I_{\text{self}} \cos^2(kx)\\
     &\quad- V_{\text{Y}} \cos^2(ky) \\
     &\quad- V_{\text{Z}} \cos^2(kz) \\
    &\quad- \sqrt{V_{\text{Xint}} V_{\text{Z}}} \cos(kz) \cos[kx + \varphi(\tau)] \\
    &\quad- I_{\text{XZ}} \sqrt{V_{\text{Xint}}(\tau) V_{\text{Z}}} \cos(kz) \cos[kx - \varphi(\tau)]~,
\end{split}
\end{equation}
where $k=2 \pi / \lambda$ and $\lambda = \SI{1064}{nm}$.
The imbalance factors $I_{\rm{self}} = 1.00(2)$ and $I_{\rm{XZ}} = 0.79(1)$ are determined through independent calibrations.
The lattice depths $[V_{\text{X}},V_{\text{Xint}},V_{\text{Y}},V_{\text{Z}}]$ used in this paper are given by $[6.04(6), 0.44(2), 22.3(1), 16.4(2)] E_R$ (Fig.~\ref{fig:2}A, Fig.~\ref{fig:3}A, B, C) and $[6.51(6), 0.39(2), 25.02(3), 17.15(9)] E_R$ (Fig.~\ref{fig:2}B), measured in units of recoil energy $E_R = h^2/2m\lambda ^2$, where $m$ is the mass of the atoms.
The values in brackets denote the standard deviations of roughly 500 and 30 measurements, respectively.

With this setup, we realise the Rice-Mele model in the tight-binding limit~\cite{walter_quantization_2023}.
The model is characterised by three numbers: the offset energy $\Delta$ between the two sites of a unit cell, the nearest-neighbour tunneling $t$ (averaged over one pump cycle), and the bond dimerisation $\delta$ which gives half the difference between the inter- and intra-dimer tunnellings.

\setlength{\tabcolsep}{0.5em}
\begin{table}[h]
\def\arraystretch{1.5}
\begin{tabular}{ ccccc }
\hline
\hline
parameter & offset & ampl. & freq. & phase offset\\
 & $B$ [Hz] & $A$ [Hz] & $\nu$ & $\kappa$ \\
\hline
$t$ & 675 & 380 & 2 & $\pi / 2$ \\ 
$\Delta$ & 0 & 1851 & 1  & $\pi$\\
$\delta$ & 0 & 989 & 1 & $\pi / 2$ \\
\hline
\end{tabular}
\caption{Time-dependent Rice-Mele parameters $t$, $\Delta$, and $\delta$  for the experimental data in Fig.~2A, Fig.~3, and Fig.~S3. The average tunnelling $t'$ (value $B$) in Fig.~2B is \SI{544}{Hz}. We describe the time-dependent parameters with a sinusoidal fit, \mbox{$B + A \sin(2\pi \nu \tau / T + \kappa)$}, where $\tau$ is time and $T$ is the pump period. The sinusoidal fit to $\Delta$ differs by at most \SI{130}{Hz} from its actual value throughout the pump cycle. The value of $t$ in Eq.~\ref{eqn:RM} corresponds to the average tunnelling (value $B$).} 
\label{tbl:rm_params}
\end{table}

The Rice-Mele parameters (Table~\ref{tbl:rm_params}) are determined by evaluating maximally localised Wannier states, which form a basis spanning the solution space of the single-particle Hamiltonian described by potential Eqn.~\ref{eqn:potential}. Through overlap integrals of these Wannier states, one can obtain the Hubbard tunnelling elements, on-site energies, and interactions $U$~\cite{uehlinger_artificial_2013}.
The relative phase $\varphi(\tau)$ between the interfering ($V_{\text{Xint}}$) and non-interfering ($V_{\text{X}}$) lattices controls their position (and amplitude) with respect to each other.
A linear ramp of $\varphi(\tau)$ modulates $\Delta$ and $\delta$ periodically in time, which can be depicted as a closed (approximately elliptical) trajectory in the $\Delta$-$\delta$ parameter space (see also Fig. \ref{fig:1}A in the main text). 
In the adiabatic limit, a Thouless pump with quantised transport can be realised. Here, the atomic displacement is given by the number of revolutions around the origin of the $\Delta$-$\delta$ plane. 


\subsection{Mapping a 1D Thouless pump to a 2D Hofstadter model with quantum Hall response}
A 1D Thouless pump with a period of $T$, as realised in our experiment, can be mapped to a 2D topological tight-binding Harper-Hofstadter-Hatsugai (HHH) model with an applied electric field $E = \frac{2 \pi \hbar}{qbT}$ where $q$ can be thought of as a fictitious charge of netural atoms and $b$ represents the lattice constant along the synthetic direction of the electric field. Due to the topological bandstructure, this electric field leads to a transverse current with current density $j_{\rm{trans}}=\frac{q}{bT}$ when considering a fully occupied band. The 2D model therefore has a quantised transverse conductance $\sigma _{\rm{trans}} =  j_{\rm{trans}}/E = \frac{q^2}{2 \pi \hbar}$ analogous to the Hall conductance in the Quantum Hall Effect (QHE).



The time-periodicity of the Hamiltonian in Eq.~\ref{eqn:RM} with ${\hat{H}(\tau) = \hat{H}(\tau + T)}$ allows us to use Floquet's theorem.
Solutions of the time-dependent Schrödinger equation
\begin{align} \label{eqn:tdse}
    i \hbar \partial _\tau \ket{\Psi(\tau)} = H(\tau) \ket{\Psi(\tau)}
\end{align}
can thus be written as 
\begin{align} \label{eqn:floquet_ansatz}
    \ket{\Psi(\tau)} = e^{-i \epsilon \tau / \hbar} \ket{u(\tau)}
\end{align}
with $\ket{u(\tau+T)}=\ket{u(\tau)}$ and $\epsilon \in \mathbb{R}$.
Due to the time-periodicity of $u(\tau)$ we expand it as a Fourier series,
\begin{align} 
    \ket{u(\tau)} = \sum _n e^{-i \omega n \tau}\ket{u_n}~,
\end{align}
where $\omega = 2\pi/T$ is the pump frequency.
The change from the time-domain into the Fourier-domain is the key ingredient to map the 1D Thouless pump to a 2D tight-binding model. 
The index $n$ is also called the photon number of the mode $\ket{u_n}$.

Using a multi-index $\alpha = (j, \sigma)$ we write the $T$-periodic 1D Hamiltonian for $U = 0$ in the Fourier-basis:
\begin{align} \label{eqn:peridic_hamiltonian}
    \hat{H}(\tau) &= \sum _{\alpha, \beta} h_{\alpha \beta} (\tau) \ket{\alpha}\bra{\beta} \\
    \nonumber &= \sum _{\alpha, \beta, m} e^{-i m\omega  \tau} h_{\alpha \beta} ^m \ket{\alpha}\bra{\beta}
\end{align}
with $h_{\alpha \beta} ^m = \frac{1}{T} \int _0 ^T e ^{i m \omega \tau} h_{\alpha \beta} (\tau)d\tau$ and $\ket{\alpha}$ corresponding to an atom localised on site $j$ with spin $\sigma$. Likewise, we use Fourier decomposition to express the solutions to Eq.~\ref{eqn:tdse} as
\begin{align} \label{eqn:eqn_floquet_ansatz_alpha}
    \ket{\Psi(\tau)} = e^{-i \epsilon \tau / \hbar} \sum _{n,\alpha} e^{-i n \omega \tau}u_{n,\alpha}\ket{\alpha}~.
\end{align}
where $u_{n,\alpha} = \braket{\alpha\vert u_n}$.
As a result, we obtain an eigenvalue equation for $u_{n,\alpha}$:

\begin{align}
    \epsilon u_{n,\alpha} = - n\hbar\omega u_{n,\alpha} + \sum _{\beta, m} h _{\alpha \beta} ^m u_{n-m,\beta} ~~\forall n,\alpha
\end{align}
which can be understood as a time independent Schrödinger equation of a 2D HHH model with a tilted potential energy along one axis. 
By explicitly evaluating the $h _{\alpha \beta} ^m$, we get

\begin{align}
    \label{eqn_hamiltonian_2D_overview}
    H_{\rm{2D}} = H_{\rm{real}} + H_{\rm{synth}} + H_{\rm{diag}} + H_{\rm{V}} + H_{\rm{tilt}},
\end{align}
with
\begin{align}
    \label{eqn:hamiltonian_2D_detail}
    &H_{\rm{real}} = -t \sum _{j,n,\sigma} (\hat{c}^\dagger_{j, n, \sigma}\hat{c}_{j+1, n, \sigma} + h.c.), \\
    \nonumber &H_{\rm{diag}} = - \frac{\delta _0}{2} \sum _{j,n,\sigma} e^{-i \pi j} (\hat{c}^\dagger_{j, n, \sigma}\hat{c}_{j+1, n+1, \sigma}+\hat{c}^\dagger_{j, n, \sigma}\hat{c}_{j+1, n-1, \sigma} + h.c.), \\
    \nonumber &H_{\rm{synth}} = - \frac{\Delta _0}{2} \sum _{j,n,\sigma} e^{-i \pi j} (i \hat{c}^\dagger_{j, n, \sigma}\hat{c}_{j, n+1, \sigma} + h.c.),\\
    \nonumber &H_{\rm{V}} = \sum _{j,n,\sigma} V(j)\hat{c}^\dagger_{j, n, \sigma}\hat{c}_{j, n, \sigma},\\
    \nonumber &H_{\rm{tilt}} = - \sum _{j,n,\sigma} \hbar\omega n \hat{c}^\dagger_{j, n, \sigma}\hat{c}_{j, n, \sigma}~.
\end{align}
$H_{\rm{real}}$ and $H_{\rm{synth}}$ describe tunneling along the real ($x$) and synthetic ($n$) dimension, respectively. 

The diagonal tunnelling terms in $H_{\rm{diag}}$ are crucial because they open a bandgap between the ground band and the first excited band, characterised by the topological Chern number $C$ which is further related to the quantised Hall conductance via $\sigma_{\rm{trans}}=\frac{q^2}{2 \pi \hbar}C$.
The terms in $H_{\rm{V}}$ describe the external potential along the real-space direction. 
$H_{\rm{tilt}}$ corresponds to a linear tilt in potential energy along the synthetic dimension which can be thought of as originating from an electric field pointing along $n$. As a consequence of this tilt, persistent Bloch oscillations occur along the synthetic dimension, while quantised Hall drifts are observed along the real dimension.


\begin{figure*}[htbp]
    \centering
    \includegraphics[width = \textwidth]{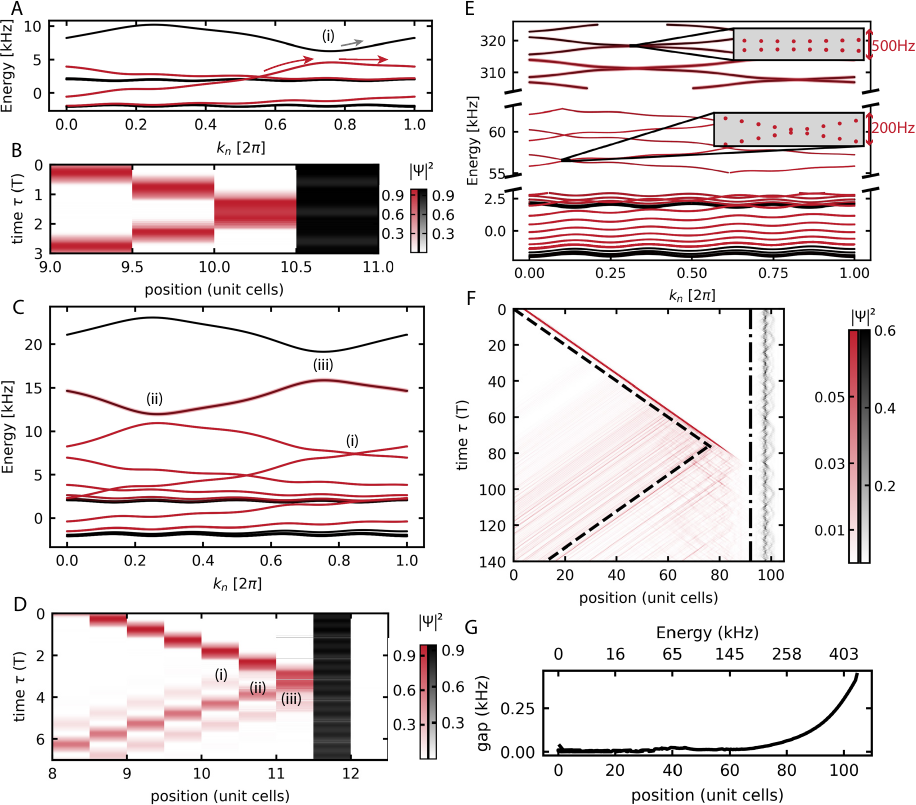}
    \caption{\textbf{Energy spectra and numerical simulations of the time evolution for confining potentials of varying steepness}. 
    The energy spectra of $H_{\rm{2D}}$ are displayed for $\kappa = 24$ (\textbf{A}), $\kappa = 10$ (\textbf{C}), and $\kappa = 2$ (\textbf{E}). 
    For clarity, localized states at the left edge are omitted. 
    Below the respective spectra, numerical simulations of single-particle evolution initiated in the topologically nontrivial (red) and trivial (black) regions are presented for $\kappa = 24$ (\textbf{B}), $\kappa = 10$ (\textbf{D}), and $\kappa = 2$ (\textbf{F}). 
    The lattice parameters, trap parameter $V_{0}$ and the pump period $T$ employed in the calculations are the same as the experiment shown in Fig.~\ref{fig:2}A.
    The dash line in (\textbf{F}) represents the linear fits to the experimentally measured atom drift, while the dash-dotted line indicates the topological boundary determined by $\Delta_{\rm{ext}}=\Delta_{0}$. 
    The topological edge states explored by the atom during the evolution are marked in red in the spectra (\textbf{A}), (\textbf{C}) and (\textbf{E}), which connect the two bands with Chern number $1$ and $-1$.
    In (\textbf{G}) we illustrate the dependence of the energy gap size in the spectrum (inset of (\textbf{E})) on the position and the energy of the corresponding eigenstates for the case of $\kappa=2$.
    }
    \label{fig:si_turnaround_spectrum}
\end{figure*}

\subsection{Edge modes and their reflection properties}
To illustrate the topological edge modes in the presence of an external potential, we evaluate the spectrum of $H_{\rm{2D}}$ in the adiabatic limit ($\omega \rightarrow 0$) with different trapping potentials
\begin{align}
\label{eqn:si_potential}
    V(j) = \frac{1}{2}m (2\pi \nu a)^2 j^{\kappa} \equiv V_{0}j^{\kappa}
\end{align}
with $m$ being the mass of $^{40}\rm{K}$, trap frequency $\nu = \SI{134}{Hz}$, lattice spacing $a=\SI{532}{nm}$, and lattice site $j$. The parameter $\kappa$ characterises the steepness of the trap and the limit $\kappa \rightarrow \infty$ corresponds to the textbook case of infinitely sharp walls~\cite{asboth_short_2016}.
Figure~\ref{fig:si_turnaround_spectrum}A, C, and E display the spectra corresponding to the cases of $\kappa=24,10,2$, respectively.
The lattice parameters and the trap parameter $V_{0}$ employed in the calculations are the same as the experiment shown in Fig.~\ref{fig:2}A.
For clarity, localized states at the left edge are omitted from the spectra.

Each spectrum is accompanied by a plot of single-particle time evolution in the respective cases of $\kappa=24,10,2$ in Fig~\ref{fig:si_turnaround_spectrum}B, D, F. These calculations, performed with the QuSpin library, employ the same lattice and trap parameters as in corresponding spectra depicted using the experimental drive period $T=\SI{4}{ms}$.
For each value of $\kappa$ we conducted two different initialisations of the time evolution.
On the one hand, a single atom is initiated at the lowest localised eigenstate (red).
On the other hand, an atom is positioned outside the topological boundary defined by $\Delta_{\rm{ext}}=\Delta_{0}$ (black).
For all values of $\kappa$, we consistently observe quantised drifts followed by reversals in the `red' cases.
On the contrary, the `black' cases exhibit stationary behavior. These numerical results confirm the existence of both topologically trivial and non-trivial regions in all three cases, thus establishing a topological boundary between topologically distinct domains.
The topological edge states are highlighted in the spectra in red. The highlighting is accomplished by analysing the atom's population on the localised eigenstates throughout the time evolution. Specifically, the brightness of the highlighting is proportional to the accumulated population over the evolution time.

For concreteness, let us now consider a very steep potential with $\kappa = 24$.
We can identify a family of topological edge states (red) in the spectrum (Fig.~\ref{fig:si_turnaround_spectrum}A), which connect the lower and the upper band (black) with Chern number $1$ and $-1$, respectively. 
Adiabatically following these edge states results in the reversal of the quantised Hall drift in real space (Fig.~\ref{fig:si_turnaround_spectrum}B).
Remarkably, during the first and third periods, the atom displays pronounced quantised drifts despite being localized outside of the bands. This indicates that the topologically nontrivial region extends up to 10 unit cells from the trap centre.
Compared to the case of a bulk-vacuum boundary, where no states exist outside the boundary, a trap with a finite $\kappa$ inevitably introduces a region with trivial topology. As a consequence, a boundary between two domains emerges, leading to the presence of topologically trivial edge states (black) above the non-trivial ones (red).
These trivial edge states are coupled to the non-trivial ones through tunnelling at the boundary, causing the hybridisation of the edge modes~\cite{mook_edge_2014}.
Consequently, persistent Bloch oscillations induced by the linear tilt in $H_{\rm{2D}}$ populate the trivial edge states through non-adiabatic transitions at the energy gap marked by (i) in Fig.~\ref{fig:si_turnaround_spectrum}A.
For the $\kappa = 24$ case this process is largely suppressed by a wide energy gap between trivial and non-trivial states.


Now, let us decrease the steepness of the confinement, taking $\kappa=10$ as an example.
In this case more than one energy gap becomes relevant for non-adiabatic dynamics, which are marked in the spectrum (Fig.~\ref{fig:si_turnaround_spectrum}C) by (i), (ii), and (iii). The size of these three gaps in absolute values are $\SI{0.08}{kHz}$, $\SI{1}{kHz}$ and $\SI{3}{kHz}$, respectively.
In general, the gap size increases as function of distance from the trap center.
As a result, the transition from topologically non-trivial to trivial edge states occurs within an extended region in space and energy, rather than suddenly.
Each gap in the spectrum gives rise to a drift reversal, denoted by (i), (ii), and (iii) in the time evolution plot (Fig.~\ref{fig:si_turnaround_spectrum}D). As before, the reflected portion of the atom corresponds to the fraction that adiabatically follows the topological edge mode, while the remaining fraction non-adiabatically crosses the gap to another edge mode.
This process agrees with the experimental observation, as shown in Fig.~\ref{fig:si_vs}B, where the reflection position becomes closer to the trap center as a function of pump period.


The experimentally relevant scenario with $\kappa=2$ ~\cite{stanescu_topological_2010,buchhold_effects_2012,wang_topological_2013,kolovsky_quantum_2014,nakajima_topological_2016} exhibits similar qualitative behavior to the $\kappa=10$ case with a greater number of localised states outside of the bands (Fig.~\ref{fig:si_turnaround_spectrum}E).
A less steep confinement causes a more gradual increase in the relevant energy gaps (Fig.~\ref{fig:si_turnaround_spectrum}E, Fig.~\ref{fig:si_turnaround_spectrum}G), widening the transition region between topologically non-trivial and trivial states. 
The corresponding drift reversal happens over a broader region in real space.
Importantly, though, the gaps between left-moving and right-moving states remain negligible for up to approximately 60 unit cells from the trap center (Fig.~\ref{fig:si_turnaround_spectrum}G).
Thus, quantised Hall drifts persist for long distances, even in a harmonic potential (Fig.~\ref{fig:si_turnaround_spectrum}F).
Based on this observation, we assert that the role of the harmonic potential in this region is primarily to localise the eigenstates, rather than altering the topological invariant associated with the winding of the Berry phase.

\begin{figure}[t!]
    \centering
    \includegraphics[width = 0.5\textwidth]{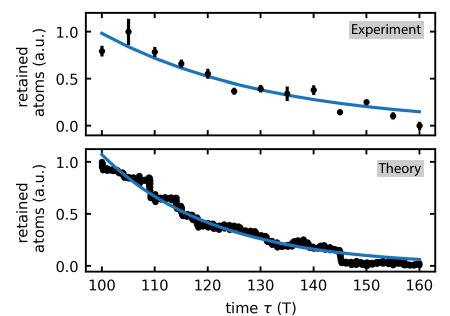}
    \caption{\textbf{Dissolving of the retention cloud at the boundary.} For both the experimental (upper panel) and numerical (lower panel) scenarios, we select a region spanning from 60 to 90 unit cells away from the trap center. We then track the temporal evolution of the total atom density within this region and fit it to a exponential decay. The fitted decay time constant is $\tau _{\rm{exp}} = 32(4)$T for the experimental result and $\tau_{\rm{sim}} = 21.1(1)$T for the numerical simulation. The initial density is normalised to 1 in both figures.}
    \label{fig:si_retention}
\end{figure}

Another feature that becomes apparent for $\kappa = 2$ is that a portion of the atoms temporarily resides at the turning point near the boundary before being reflected, as shown in Fig.~\ref{fig:si_turnaround_spectrum}F.
This retention behavior is a result of non-adiabatic transitions occurring over multiple pump periods.
In order to quantify this phenomenon, we evaluate the atom density in the edge region (60 to 90 unit cells from the centre) both in theory (Fig.~\ref{fig:si_turnaround_spectrum}F) and in the experiment (Fig.~\ref{fig:2}A), as function of time.
This data is shown in Fig.~\ref{fig:si_retention} and the timescales for atom retention at the edge roughly agree within theory and experiment.


\begin{figure}[t!]
    \centering
    \includegraphics[width = 0.5\textwidth]{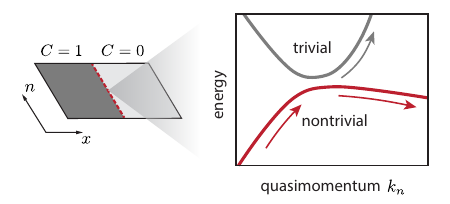}
    \caption{\textbf{Hybridisation of the edge modes at the topological interface.} A segment of the energy spectrum showcasing the presence of both topologically trivial and non-trivial edge modes at the interface between two regions with distinct Chern numbers. 
    Tunnelling-induced couplings at the topological interface can hybridise the trivial and non-trivial edge modes, while keeping the total count of gapless edge modes at the boundary unchanged.
    Bloch oscillations along $k_n$ lead to non-adiabatic transfers from the topologically non-trivial to the trivial edge modes.
    The population of trivial edge modes explains the transient retention of the atoms at the boundary before subsequent Bloch oscillations transferring atoms back into the topological domain.}
    \label{fig:si_retention_theory}
\end{figure}

The hybridization of the edge modes, as previously mentioned, is a universal phenomenon occurring at the interface between topologically trivial and nontrivial regions.
As an example, let us consider a sharp domain wall between $C=1$ and $C=0$ (Fig.~\ref{fig:si_retention_theory}).
According to the bulk-edge correspondence, the topologically nontrivial region contributes exactly one gapless mode whereas the trivial region can contribute gapped edge modes.
Depending on the specific structure of the system, the trivial and nontrivial edge modes can exhibit a crossing, a touching, or remain separated by an energy gap.
Tunnelling-induced couplings at the topological interface can hybridise the trivial and non-trivial edge modes, leading to either the opening of an energy gap or the modification of the size of an existing gap.
The system presented in this work aligns more closely with the latter scenario while the smooth confinement leads to a more complex level structure (Fig.~\ref{fig:si_turnaround_spectrum}).
It is noteworthy that, in the context of such hybridization, the total number of gapless edge modes remains unaltered, in accordance with the bulk-edge correspondence.
If the gap between the topological and the trivial edge modes is on the order of the pump frequency $2\pi/T$, Bloch oscillations along $k_n$ can lead to non-adiabatic transfers between these edge modes, resulting in temporary population of the trivial edge states before subsequent Bloch oscillations transfering them back to the topological ones.
In real space, this manifests as atoms transiently residing at the boundary, then drifting back.

\subsection{Staggered potential}
An alternative approach to identify the topological boundary in our system makes use of a staggered potential.
First, we consider a potential with uniform staggering, given by $V_{c}(j) = V (-1)^j$, where $j$ indexes the lattice-site and $2V$ corresponds to the energy difference between adjacent sites. Adding such a potential to the Rice-Mele Hamiltonian (Eq.~\ref{eqn:RM}) changes its trajectory in the $\Delta$-$\delta$ plane. The onsite energy in such a system is given by $(\Delta (\tau) + V)(-1)^j$, which ranges from $-\Delta + V$ to $\Delta + V$. The tunnellings are unchanged. Therefore, the trajectory remains circular and it is simply shifted upwards by an amount $V$.

\begin{figure}[t!]
    \centering
    \includegraphics[width = 0.48\textwidth]{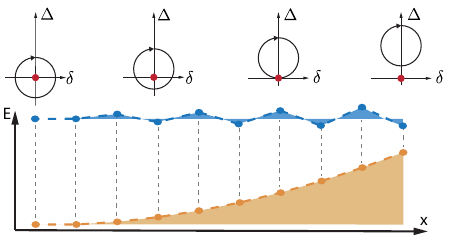}  
    \caption{\textbf{Linearly increasing staggered potential.} To elucidate the topology in our system, a linearly increasing staggered potential is considered (blue): $V(j) = j V_0(-1)^j$ with $V_0 = 1/2\times m(2\pi \nu a)^2$, as before. It is chosen such that the local tilt always equals the tilt from the harmonic potential (orange), but alternates in sign. The staggered potential allows a simple pictorial representation of the emergence of the topological boundary. In a local density approximation the pump trajectory is linearly shifted upwards in the $\Delta$-$\delta$ plane~\cite{nakajima_topological_2016}, as depicted in the upper part of the figure. As soon as the trajectory ceases to enclose the critical point, a topological--trivial boundary develops.}
    \label{fig:si_staggered}
\end{figure}

A topological boundary emerges for a linearly increasing staggered potential, given by $V(j) = j V_0(-1)^j$, with $V_0 = \frac{1}{2}m(2\pi \nu a)^2$ as before. $V(j)$ is chosen such that it has the same local tilt as the harmonic trap in the experiment. Within the local density approximation we assign a $\Delta$-$\delta$ trajectory locally to each unit cell.
The trajectories are thus linearly shifted upwards as function of $j$ (Fig.~\ref{fig:si_staggered}), describing a change of topology in real space.
We expect the local density approximation to be valid since the atomic eigenstates in the experiment are strongly localised.

\begin{figure}[h]
    \centering
    \includegraphics[width = 0.5\textwidth]{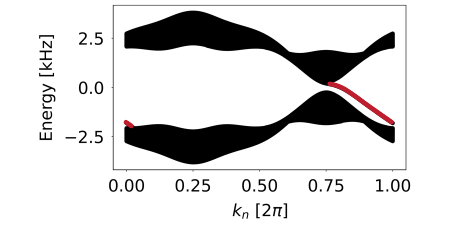}
    \caption{\textbf{Edge state spectrum for a smoothly increasing staggered potential.} The spectrum shows two bands, corresponding to Chern numbers $C= 1$ and $C = -1$, respectively. A single gapless edge mode, marked in red, connects the two bands of opposite Chern number. The lattice parameters and $V_{0}$ for this calculation are the same as in the experiment shown in Fig.~\ref{fig:2}A.}
    \label{fig:si_staggered_spectrum}
\end{figure}

Fig.~\ref{fig:si_staggered_spectrum} shows the spectrum for the linearly increasing staggered potential, which allows for a straightforward identification of the gapless edge mode (red line) with a smooth boundary. The states corresponding to this gapless edge mode are localised around the topological boundary.
Similar models with linearly increasing staggered potential have been studied in refs.~\cite{stanescu_topological_2010,goldman_creating_2016,irsigler_interacting_2019}.

\subsection{Local Chern marker}
The mathematical formulation of the Chern number as a topological invariant requires translational invariance, which does not apply to realistic experiments.
Instead, we use a local quantity, known as Chern marker~\cite{bianco_mapping_2011,chalopin_probing_2020}. The local Chern marker depends on the real-space position and it is defined by:
\begin{align}
    \label{eqn:si_chern_marker_definition}
    c(r_{\gamma}) = - \frac{4 \pi}{A_c} Im \sum _{s=A,B} \bra{r_{\gamma _s}} \hat{P} \hat{x} \hat{Q} \hat{y} \hat{P} \ket{r_ {\gamma _s}},
\end{align}
where $r_{\gamma}$ is the position of the unit cell $\gamma$ with sub-lattice-sites at positions $r_{\gamma_A}$ and $r_{\gamma_B}$, $\ket{r_ {\gamma _s}} = c^{\dagger}_{\gamma _s}\ket{0}$ is the state localised on the corresponding lattice site , $A_c$ is the area of a real-space unit cell, $\hat{Q} = 1 - \hat{P}$ and $\hat{P}$ is the projector onto the ground band.
Defining $\hat{P}$ is not unambiguously possible in our system (Eq.~\ref{eqn_hamiltonian_2D_overview}) because of the energy shift from the harmonic confinement.
Instead, we use a linearly increasing staggered potential, as described in the previous paragraph.
This model leaves the bands intact and a ground band can be unambiguously defined.
Experimentally, a local probe of the band topology is the velocity of the Hall drift, plotted in Fig.~\ref{fig:2}A.
Theory and experiment agree approximately with one another.
The local velocity is extracted from the atomic positions by fitting linear functions to groups of three adjacent data-points in ten pump cycles. The resulting velocities are plotted against position and smoothed through a running average of width three (ten cycles).

\begin{figure}[htbp]
    \centering
    \includegraphics[width = 0.48\textwidth]{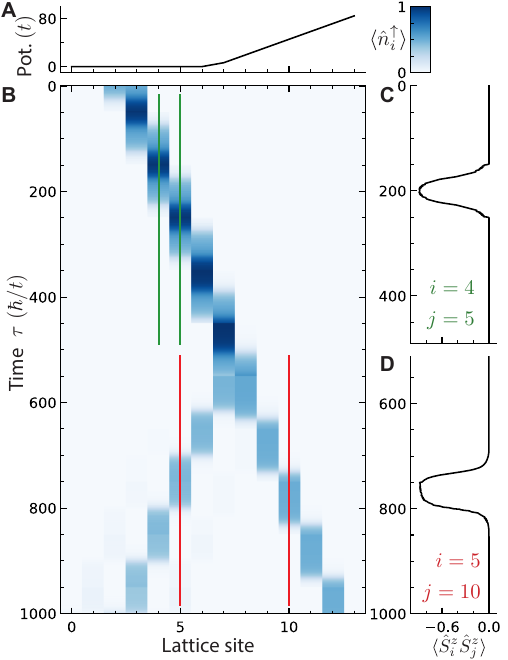}  
    \caption{\textbf{Numerical evidence for splitting of singlets.} (\textbf{A}) suddenly changing potential gradient in the middle of the lattice.
    (\textbf{B}) the time-evolved density $\braket{\hat{n}_i^\uparrow}$ for all lattice sites.
    Initially, the singlet pair moves as a whole, periodically converted between singlet $(\ket{\uparrow,\downarrow} - \ket{\downarrow,\uparrow})/2$ at $\tau \in \{0,T/2,T,\dots\}$ and double occupancy at $\tau \in \{T/4,3T/4,\dots\}$.
    After the splitting has taken place, the singlet is separated in space and each constituent atom is pumped quantisedly in opposite directions.
    The spin correlators between sites 4 and 5 (\textbf{C}, green lines in \textbf{B}), as well as 5 and 10 (\textbf{D}, red lines in \textbf{B}) are plotted as function of time (vertical axes).
    The pump period for this figure is $200~\hbar/t$.
    }
    \label{fig:si_singlet}
\end{figure}

\subsection{Splitting of singlet pairs}

To describe the splitting process of spin singlet due to an increasing gradient, we perform exact diagonalisation calculations in a small system.
We simplify the situation by considering the time evolution of a single singlet pair, $(\ket{\uparrow,\downarrow} - \ket{\downarrow,\uparrow})/2$, initially localised on sites $2$ and $3$ of a small lattice (Fig.~\ref{fig:si_singlet}).
The time evolution is performed with a pump period of $200~\hbar/t$ and $U/t = 10$, using the QuSpin library.
Similar to ref.~\cite{qian_quantum_2011}, we use a rectangular pumping profile which spends 99 per cent of the pump period in linearly ramping $\Delta$ from $-\Delta_{\text{max}}$ to $+\Delta_{\text{max}}$ and vice versa with $\Delta_{\text{max}} = 10t$. The remaining 1 per cent of the cycle is spent to change the dimerisation (also linearly) from $-\delta_{\text{max}}$ to $+\delta_{\text{max}}$ and vice versa ($\delta_{\text{max}} = t$). 
In this manner, the delocalisation of atoms due to tunnelling is suppressed and we can follow a single pair of atoms along the pump.
The increasing gradient is modelled by a linear potential which suddenly starts in the middle of the lattice with a strength of $13t$ per site (Fig.~\ref{fig:si_singlet}A).
This potential exactly reproduces the behaviour illustrated in Fig.~\ref{fig:3}E.
While on the left of the lattice the pump trajectory encloses both singularities, the trajectory only encloses one of the two singularities on the right.
It is evident from the density $\braket{\hat{n}_i^\uparrow}$ (Fig.~\ref{fig:si_singlet}B) that the singlet pair moves as a whole for the first two pump cycles.
After the atoms start to experience the gradient, the pair of atoms splits into two parts, each pumping quantisedly in opposite directions.
The $\braket{\hat{n}_i^\downarrow}$ density evolves identically to $\braket{\hat{n}_i^\uparrow}$.
We evaluate the spin correlator
\begin{equation*}
\braket{\hat{S}^z_i\hat{S}^z_j} = \langle\Psi(\tau)| (\hat n_{\uparrow,i}-\hat n_{\downarrow,i}) (\hat n_{\uparrow,j}-\hat n_{\downarrow,j})|\Psi(\tau)\rangle\\
\end{equation*}
as function of time for $(i,j) = \{(4,5),(5,10)\}$, where $\ket{\Psi(\tau)}$ is time-evolved many-body state.
The results are plotted in Fig.~\ref{fig:si_singlet}C,D.
The correlator $\braket{\hat{S}^z_4\hat{S}^z_5}$ (Fig.~\ref{fig:si_singlet}C) starts off at zero because no atoms are initially present at sites $4,5$.
Once the singlet pair comes past the sites $4,5$ we observe a strong anti-ferromagnetic signal, which demonstrates the pumping of a singlet as a whole.
Interestingly, when considering the correlations between site 5 and 10 (Fig.~\ref{fig:si_singlet}D), the anti-ferromagnetic signal is almost equally strong, albeit separated by five lattice sites.
We conclude that singlet pairs remain correlated after the gradient-induced splitting process.

\end{document}